\documentclass[10pt,prl,aps,floatfix,superscriptaddress,footinbib,longbibliography,reprint]{revtex4-2}

\usepackage{silence}
\usepackage{graphicx,mwe} 
\usepackage{hyperref}
\usepackage[dvipsnames]{xcolor}
\usepackage{amsmath,braket}
\usepackage[normalem]{ulem}
\DeclareRobustCommand{\coloneqq}{\mathrel{\mathop:}=}
\usepackage{acronym}
\newacro{DMFT}[DMFT]{dynamical mean field theory}
\newacro{ED}[ED]{Exact Diagonalization}
\newacro{MC}[MC]{Monte Carlo}
\newacro{QMC}[QMC]{quantum Monte Carlo}
\newacro{SEET}[SEET]{self-energy embedding theory}
\newacro{SOEHYB}[SOEHYB]{sum-of-exponentials hybridization expansion solver}
\newacro{CTHYB}[CTHYB]{continuous-time hybridization-expansion quantum Monte Carlo}
\newacro{ESPRIT}[ESPRIT]{Estimation of Signal Parameters via Rotational Invariance Techniques}

\def\umphys{
    Department of Physics, University of Michigan,
    Ann Arbor, Michigan 48109, USA
}
\def\umchem{
Department of Chemistry, University of Michigan, Ann Arbor, Michigan 48109, USA
}
\def\uwphys{
Institute of Theoretical Physics, Faculty of Physics, University of Warsaw, Warsaw, Poland
}

\def\cas{
Beijing National Laboratory for Condensed Matter Physics and Institute of Physics, \\Chinese Academy of Sciences, Beijing 100190, China
}

\begin{document}

\title{Hybrid Hamiltonian-diagrammatic quantum impurity solver}
\author{Yang Yu}
\email{umyangyu@umich.edu}
\affiliation{\umphys}
\author{Gaurav Harsha}
\affiliation{\umchem}
\author{Lei Zhang}
\affiliation{\umphys}
\author{Agnieszka Jażdżewska}
\affiliation{\uwphys}
\author{Dominika Zgid}
\affiliation{\umphys}
\affiliation{\umchem}
\affiliation{\uwphys}
\author{Xinyang Dong}
\email{dongxy@iphy.ac.cn}
\affiliation{\cas}
\author{Emanuel Gull}
\email{egull@umich.edu}
\affiliation{\umphys}
\affiliation{\uwphys}

\begin{abstract}
Quantum impurity models, which describe the coupling between interacting orbitals and a non-interacting bath, play a central role in the physics of strongly correlated electron systems. 
Solving a quantum impurity model in general requires the use of non-perturbative numerical methods. 
Hamiltonian-based approaches, which rely on an explicit bath discretization,  are typically limited to a small number of bath sites or small entanglement, and diagrammatic methods suffer from sign problems, slow convergence, or diagram truncation approximations. 
Here we show that these two classes of methods can be combined: augmenting diagrammatic methods with a small auxiliary bath can reduce the residual problem to a regime where low-order perturbation theory is highly accurate and rapidly converging. 
In a simple benchmark, the precision of the hybrid approach surpasses bold-line calculations by several orders of magnitude; for a strongly interacting two-orbital model with a severe sign problem, convergence is achieved at three orders of magnitude lower computational cost than competing methods; and convergence to the unknown exact result is rapidly accelerated in a difficult realistic problem. Our results establish a practical route to high-precision quantum impurity solutions in correlated quantum systems.
\end{abstract}

\maketitle

Quantum impurity models describe a small set of interacting orbitals coupled 
to a continuous non-interacting bath. Originally introduced to model magnetic impurities embedded in 
a metallic host \cite{andersonLocalizedMagneticStates1961}, they also arise in physical contexts ranging from  
quantum dots and molecular conductance \cite{hansonSpinsFewelectronQuantum2007} to atoms adsorbed on surfaces \cite{langrethDerivationMasterEquation1991}. Most      
importantly, they appear as auxiliary problems in embedding theories such as the \ac{DMFT} \cite{georgesDynamicalMeanfieldTheory1996,kotliarElectronicStructureCalculations2006} and the \ac{SEET} \cite{zgidFiniteTemperatureQuantum2017,rusakovSelfEnergyEmbeddingTheory2019,Tran2017GeneralizedSEET}.   
In general, their solution requires powerful non-perturbative numerical methods.

One class of  solvers relies on casting the impurity action into an explicit Hamiltonian form. This    
requires approximating the continuous hybridization function by discretizing it into a finite set of bath energy levels. For a small
number of bath levels, exact diagonalization \cite{caffarelExactDiagonalizationApproach1994} is used to obtain the solution of this discretized system.
Quantum chemistry methods, such as configuration interaction                
\cite{Zgid2011DMFT,zgidTruncatedConfigurationInteraction2012} and coupled cluster theory \cite{sheeCoupledClusterImpurity2019,zhuCoupledclusterImpuritySolvers2019}, 
as well as density matrix renormalization group methods \cite{nishimotoDensitymatrixRenormalizationGroup2004,garciaDynamicalMeanField2004} (including a
new generation of complex-time solvers \cite{caoDynamicalCorrelationFunctions2024,grundnerComplexTimeEvolution2024,yuMultiorbitalDynamicalMeanfield2026}) extend this to a much    
larger number of bath states. Efficient generalizations to real frequencies employing very many bath sites also exist \cite{Lu2014,luExactDiagonalizationImpurity2017}.

Another class of methods performs a diagrammatic expansion of the impurity action, typically with   
\ac{MC} sampling. Continuous-time \ac{QMC} methods                              
\cite{rubtsovContinuoustimeQuantumMonte2005,wernerContinuoustimeSolverQuantum2006,gullContinuoustimeAuxiliaryfieldMonte2008,gullBoldlineDiagrammaticMonte2010,gullContinuoustimeMonteCarlo2011,eidelsteinMultiorbitalQuantumImpurity2020} are exact up to their stochastic      
error, which can become prohibitively large in the presence of a sign problem. Even in its absence,  
convergence to the exact result is limited by the slow $\sim N^{-1/2}$ scaling of \ac{MC} errors 
with the number of samples $N$. One may instead truncate the diagrammatic series at low order and supplement it with a self-consistent partial summation
to obtain an efficient though uncontrolled approximation \cite{keiterPerturbationTechniqueAnderson1970,pruschkeAndersonModelFinite1989}. Recent advances in diagram compression with tensor networks \cite{nunezfernandezLearningFeynmanDiagrams2022,yuInchwormTensorTrain2025,erpenbeckTensorTrainContinuous2023} and separable representations \cite{kayeDecomposingImaginaryTimeFeynman2024,kayeCppdlrImaginaryTime2024,huangAutomatedEvaluationImaginary2025} have far extended the parameter space accessible with diagrammatic techniques.

While both Hamiltonian and diagrammatic methods converge to the exact result in principle, their
convergence in practical applications often leaves much to be desired. This is especially true for   
quantum impurity problems arising in realistic systems \cite{zgidFiniteTemperatureQuantum2017}, which feature complex impurity interactions, 
off-diagonal hybridization functions, and strong entanglement and correlation at low temperature. For such systems, neither class of methods reliably reaches the exact solution, and 
the development of accurate, efficient quantum impurity solvers remains a critical open problem.     

This work shows that combining Hamiltonian and
diagrammatic methods leads to a surprising acceleration of convergence. We augment the impurity space
with a small set of auxiliary bath levels - which we term \emph{counterterms} - and solve the       
resulting enlarged local problem non-perturbatively, capturing the dominant contributions of the bath
in a few discrete modes. The difference between the full hybridization and its counterterm          
approximation defines a small residual hybridization, which is then treated diagrammatically. Because
the counterterms are chosen to minimize this residual, the perturbative expansion converges 
at low order.
                               
We demonstrate that the accuracy of such a solver far exceeds existing methods for an exactly solvable but (in the hybridization expansion framework) non-trivial benchmark — the
spinless non-interacting Anderson impurity model. For a typical challenging case, a strongly         
interacting two-orbital impurity model with a severe sign problem, we find rapid and systematic
convergence along with a substantial computational speedup. Finally, for a realistic embedding problem, where no accurate reference solution exists, 
we show that the results systematically improve with increasing counterterms and expansion order, although achieving full convergence remains challenging.
Combining counterterms with higher-order diagrammatics thus provides a systematic path
toward high-precision solutions of quantum impurity models of this type.

We consider the impurity action \cite{gullContinuoustimeMonteCarlo2011} 
\begin{align}
    \mathcal{S}_{\mathrm{imp}}[c^{*},c] 
    =  \mathcal{S}_{\mathrm{loc}}[c^{*},c] 
      + \mathcal{S}_{\mathrm{hyb}}[c^*,c],
\end{align}
where the impurity degrees of freedom $\{c^*\}$ and $\{c\}$ interact according to the local action \(\mathcal{S}_{\mathrm{loc}}[c^{*},c]\)  and hybridize with a continuous bath described by
\begin{align}
    \mathcal{S}_{\mathrm{hyb}}[c^*,c]= \iint_{[0,\beta]^2} \hspace{-0.5cm}\mathrm{d}\tau\mathrm{d}\tau' \sum_{\alpha\alpha'} c^{*}_{\alpha}(\tau)\Delta_{\alpha\alpha'}(\tau-\tau')c_{\alpha'}(\tau'), \nonumber
\end{align}
where  $\Delta$ is a hybridization function, $\alpha$ a composite spin-orbital index, and $\beta$ the inverse temperature.

As was previously explored in the context of diagrammatic \ac{MC} \cite{polletRegularizationDiagrammaticSeries2010,rossiShiftedactionExpansionApplicability2016,rossiDeterminantDiagrammaticMonte2017,kimHomotopicActionPathway2021,wangVariationalDiagrammaticMonte2025,liDiagrammaticMonteCarlo2020,liInteractionexpansionInchwormMonte2022} and dual fermion theory \cite{rubtsovDualFermionApproach2008,rohringerDiagrammaticRoutesNonlocal2018}, one may change the relative size of these terms by adding an arbitrary contribution to $\mathcal{S}_\mathrm{loc}$ and subtracting the identical term from $\mathcal{S}_{\mathrm{hyb}}$.
Here, we propose to use 
\begin{align}\label{eqn:Delta_CT}
&\mathcal{S}_{\mathrm{CT}}
    = \iint_{[0,\beta]^2} \hspace{-0.5cm} \mathrm{d}\tau \mathrm{d}\tau' \sum_{\alpha\alpha'} c^{*}_{\alpha}(\tau)\Delta^{\mathrm{CT}}_{\alpha\alpha'}(\tau-\tau')c_{\alpha'}(\tau'),\\
    &\Delta^{\mathrm{CT}}_{\alpha\alpha'}(\tau-\tau')
    = \sum_{j=1}^{n_\mathrm{CT}} 
      V_{\alpha j}\,V_{\alpha' j}^{*}\,
      g_{j}(\tau-\tau'),
\end{align}
where $j=1,\dots, n_{\mathrm{CT}}$ denotes a finite, presumably small number of \emph{counterterms}, $V_{\alpha j}$ is the coupling between the impurity orbital $\alpha$ and the auxiliary orbital $j$, and each \(g_j\) is chosen as the Green’s function of a non-interacting fermionic level with energy \(\epsilon_j\).

A Hubbard–Stratonovich transformation (see Supplemental Material~\cite{supplement_counterterms}\nocite{broydenConvergenceClassDoublerank1970,powellBOBYQAAlgorithmBound2009,royESPRITestimationSignalParameters1989,huaMatrixPencilMethod1990,sarkarUsingMatrixPencil1995,feiAnalyticalContinuationMatrixvalued2021,nakatsukasaAAAAlgorithmRational2018}) results in a revised impurity action 
\begin{align}\label{eqn:renormalized_action}
    \mathcal{S}^{\mathrm{R}}_{\mathrm{imp}}[c^{*},c,f^{*},f]
    =
    \mathcal S^{\mathrm{R}}_{\mathrm{loc}}[c^*,c,f^{*},f] 
    +
    \mathcal S^{\mathrm{R}}_{\mathrm{hyb}}[c^*,c],
\end{align}
which explicitly includes these counterterms as auxiliary fermions $\{f^*\}$ and $\{f\}$. $\mathcal{S}^{\mathrm{R}}_{\mathrm{imp}}[c^{*},c,f^{*},f]$ is equivalent to \(\mathcal{S}_{\mathrm{imp}}[c^{*},c]\) in the sense that the correlation functions for the original fermions \(\{c^{*}\}\) and \(\{c\}\) are identical to those of $\mathcal{S}_{\mathrm{imp}}$. Explicitly, the local and hybridization parts are
\begin{align}
    \mathcal{S}^{\mathrm{R}}_{\mathrm{loc}}[c^*\!\!,c,f^{*}\!\!,f]
    &= \nonumber
    \mathcal{S}_{\mathrm{loc}}[c^*\!\!,c] 
      + \int_{0}^{\beta}\!\!\!\!\mathrm{d}\tau\sum_{j}
      f^{*}_{j}(\tau)(\partial_{\tau}+\epsilon_{j})f_{j}(\tau) \\
      + \int_{0}^{\beta}\mathrm{d}\tau\sum_{\alpha j}&\left[
      V^{*}_{\alpha j}f^{*}_{j}(\tau)c_{\alpha}(\tau)
      +V_{\alpha j}c_{\alpha}^{*}(\tau)f_{j}(\tau)\right
      ], \nonumber \\
    \mathcal{S}^{\mathrm{R}}_{\mathrm{hyb}}[c^*\!\!,c]
    &=\!\! \iint_{[0,\beta]^2}\!\!\!\!\!\!\!\!\!\! \mathrm{d}\tau \mathrm{d}\tau' \sum_{\alpha\gamma} c^{*}_{\alpha}(\tau)\Delta^{\mathrm{R}}_{\alpha\gamma}(\tau\!-\!\tau')c_{\gamma}(\tau'), \nonumber
\end{align}
with residual hybridization function $\Delta^{\mathrm{R}}\coloneqq \Delta - \Delta^{\mathrm{CT}}$.

Hamiltonian-based quantum impurity solvers, such as \ac{ED} \cite{caffarelExactDiagonalizationApproach1994}, configuration-interaction \cite{Zgid2011DMFT,zgidTruncatedConfigurationInteraction2012}, or density matrix renormalization group theory \cite{nishimotoDensitymatrixRenormalizationGroup2004,garciaDynamicalMeanField2004}, solve for $\mathcal{S}^{\mathrm{R}}_{\mathrm{loc}}[c^*,c,f^{*},f]$ while 
neglecting $\mathcal{S}^{\mathrm{R}}_{\mathrm{hyb}}[c^*,c]$. They approach the exact limit by increasing the number of counterterms (bath sites) and thereby converging $\Delta^\mathrm{R}$ to zero. 
In contrast, continuous-time \ac{QMC} \cite{wernerContinuoustimeSolverQuantum2006,gullContinuoustimeMonteCarlo2011} treats $\mathcal{S}_{\mathrm{loc}}[c^{*},c] $ exactly and performs a perturbative expansion in terms of $\mathcal{S}_{\mathrm{hyb}}[c^*,c]$ to all orders, without relying on counterterms.

The choice of counterterms presents a difficult optimization problem that has been discussed extensively in this context and is known as `bath parameterization', with methods ranging from non-linear fits \cite{caffarelExactDiagonalizationApproach1994} to semidefinite relaxation \cite{mejuto-zaeraEfficientHybridizationFitting2020}. Recent developments include methods based on pole estimation with projection and semi-definite relaxation \cite{huangRobustAnalyticContinuation2023,huangAutomatedEvaluationImaginary2025} or with Prony approximants \cite{zhangMinimalPoleRepresentation2024a,zhangMinimalPoleRepresentation2024,zhangMinimalPoleRepresentation2025} obtained via signal processing techniques \cite{yingPoleRecoveryNoisy2022,yingAnalyticContinuationLimited2022}, which show systematic, often exponential, convergence as a function of the number of counterterms. Whereas the standard approaches discussed above perform the fitting procedure in frequency space,  we carry out an imaginary-time fit in this work using the matrix-valued \ac{ESPRIT} algorithm of Ref.~\cite{zhangMinimalPoleRepresentation2024} (see Supplemental Material~\cite{supplement_counterterms} for details). This strategy is closely related to the truncated Hankel correlator method of Ref.~\cite{ostmeyerTruncatedHankelCorrelator2025}.

Bare hybridization expansions \cite{wernerContinuoustimeSolverQuantum2006} solve Eq.~\eqref{eqn:renormalized_action} by expanding $\mathrm{e}^{-\mathcal{S}_{\mathrm{hyb}}^{\mathrm{R}}}$ into a power series. They are rapidly convergent when $\mathcal{S}^{\mathrm{R}}_{\mathrm{hyb}}$ is small. The impurity partition function can be written as
\begin{align}\label{eqn:bare}
    Z^{\mathrm{R}}_{\mathrm{imp}}
    =\sum\limits_{n=0}^{\infty} \sum\limits_{\substack{1 \cdots n \\ 1'\cdots n'}} \int\mathcal{D}[d^{*}, d]\frac{\mathrm{e}^{- \mathcal{S}^{\mathrm{R}}_{\mathrm{loc}}} }{(n!)^{2}}  d_{n}d^{*}_{n'}  \cdots
    d_{1}d_{1'}^{*} \operatorname{det} \mathbf{\Delta}^{\mathrm{R}},
\nonumber \end{align}
where $\{d^*\}$ and $\{d\}$ collect both original fermions and auxiliary fermions, and the subscripts represent composite indices encompassing all fermionic degrees of freedom. The determinant $\operatorname{det} \mathbf{\Delta^{\mathrm{R}}} \coloneqq \sum\limits_{P\in S_{n}} (-1)^{n_{P}} \Delta^{\mathrm{R}}_{1'P(1)} \cdots \Delta^{\mathrm{R}}_{n'P(n)}$, where $P$ ranges over all permutations $S_{n}$ of the set $\left\{1,\cdots,n\right\}$ and $n_{P}$ is the number of exchanges in $P$. An analogous expression for impurity Green's functions \cite{wernerContinuoustimeSolverQuantum2006} can be obtained by  inserting additional original fermions into the expression. Here, we use bare expansions to maximum order $k=1$ and $2$.

The  exactly solvable non-interacting Anderson impurity model presents a stringent test case for hybridization expansion methods \cite{wernerContinuoustimeSolverQuantum2006}, since interactions lower the required expansion order of the perturbation theory around the atomic limit \cite{wernerContinuoustimeSolverQuantum2006,gullPerformanceAnalysisContinuoustime2007}. We therefore first examine a spinless fermion site coupled to bath characterized by a semicircular density of states $\Gamma(\omega) = \sqrt{4t^{2}-\omega^{2}}/(2\pi t^{2}),$ $-2t \leq \omega \leq 2t,$ (hopping $t$ sets energy units) and compute the Green's function $G(\tau)$.

\begin{figure}[tb]
    \centering
    \includegraphics[width=1\linewidth]{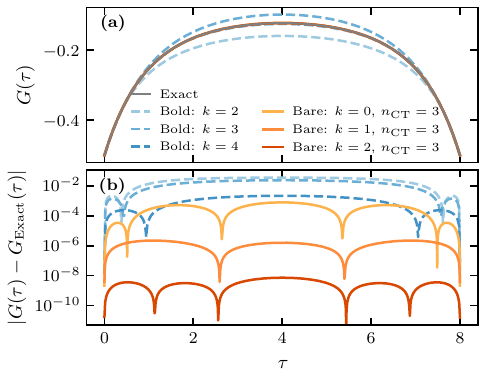}
    \caption{Spinless, non-interacting impurity model coupled to a semi-circular bath at $\beta t = 8$. Top: Green's function. Bottom: deviation from the exact result. $n_{\mathrm{CT}}$: number of counterterms,  $k$: maximum  expansion order. Bold-line calculated using \texttt{triqs\_xca}~\cite{kayeDecomposingImaginaryTimeFeynman2024,huangAutomatedEvaluationImaginary2025,kayeCppdlrImaginaryTime2024,kayeDiscreteLehmannRepresentation2022}.}
    \label{fig:spinless}
\end{figure}
The top panel of Fig.~\ref{fig:spinless} presents $G(\tau)$. The exact solution is shown in gray, along with the bold-line \cite{huangAutomatedEvaluationImaginary2025} solution at order $k=2,3,$ and $4$ (blue dashed lines). Also shown is the result for $n_\mathrm{CT}=3$ at bare perturbation  order $k=0, 1, $ and $2$ (orange and red lines).
The lower panel shows the deviation from the exact solution on a logarithmic axis. While bold perturbation theory at fourth order reaches close to $10^{-4}$, the counterterm formalism decreases by around two orders of magnitude at every order of the bare series, reaching $\varepsilon = 10^{-9}$ at $k=2$. 

\begin{figure}[tb]
    \centering
    \includegraphics[width=1\linewidth]{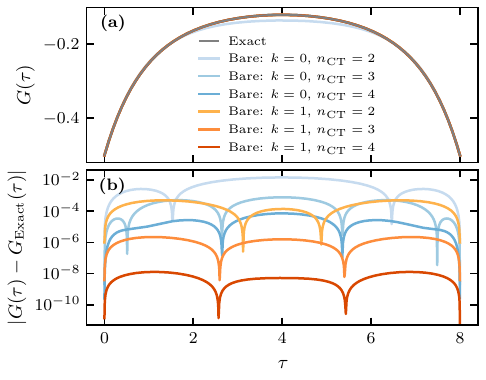}
    \caption{Spinless, non-interacting impurity model coupled to a semi-circular bath at $\beta t = 8$. Top: Green's function. Bottom: deviation from the exact result. $n_{\mathrm{CT}}$: number of counterterms. $k$:  maximum  expansion order. Bold-line calculated using \texttt{triqs\_xca}~\cite{kayeDecomposingImaginaryTimeFeynman2024,huangAutomatedEvaluationImaginary2025,kayeCppdlrImaginaryTime2024,kayeDiscreteLehmannRepresentation2022}.}
    \label{fig:spinless_nCT}
\end{figure}
Fig.~\ref{fig:spinless_nCT} shows the same system as Fig.~\ref{fig:spinless} and examines convergence as a function of the number of counterterms. $k=0$ corresponds to a Hamiltonian system. $k=1$ denotes the lowest order perturbation theory. Shown is rapid convergence to the exact solution with $n_\textrm{CT}$, with every additional counterterm improving precision by around two orders of magnitude. The solution with $n_\textrm{CT}=3$, $k=2$ is roughly comparable in accuracy to the one with $n_\textrm{CT}=4$, $k=1$, showing that both increasing the number of counterterms (at exponential cost in the size of the local Hilbert space) and increasing the diagram order (at factorial cost in the number of diagrams) yield comparable convergence to the exact result. In practice, the preferable choice will depend both on details of the impurity Hamiltonian (such as point-group symmetries) and on details of the hybridization terms (such as the rank of the decomposition in \cite{kayeLibdlrEfficientImaginary2022}).

\begin{figure}[tb]
    \centering
    \includegraphics[width=1\linewidth]{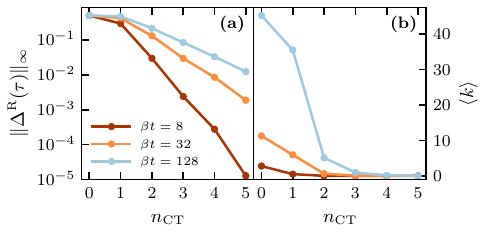}
    \caption{Spinless, non-interacting impurity model coupled to a semi-circular bath at $\beta t = 8,32,128$. Left: size of residual hybridization function. Right: average expansion order. $n_{\mathrm{CT}}$ denotes the number of counterterms.}
    \label{fig:spinless_combined}
\end{figure}
The rapid convergence to the exact result can best be understood by analyzing the `average' expansion order \cite{rubtsovContinuoustimeQuantumMonte2005} of the bare hybridization expansion, which is given by 
$\braket{k} = -\sum_{12} \Delta^{\mathrm{R}}_{12}  G_{21}$ \cite{hauleQuantumMonteCarlo2007}.
While the impurity Green’s function $G$ is problem-specific (and exactly known in the current example),  the hybridization strength $\Delta^\mathrm{R}$ depends on the number of counterterms. The left panel of Fig.~\ref{fig:spinless_combined} shows the magnitude of this residual hybridization $\Delta^{\mathrm{R}}$ for the example of Fig.~\ref{fig:spinless}, the right one the average expansion order $\braket{k}$ for different temperatures. As more counterterms are added, the magnitude of $\Delta^{\mathrm{R}}$ rapidly decreases to the weak coupling limit, with a corresponding decrease in $\braket{k}$. A detailed analysis of the resulting Green's functions as a function of $\beta$, $n_\text{CT}$, and $k$ is presented in the Supplemental Material~\cite{supplement_counterterms}.

\begin{figure}[tb]
    \centering
    \includegraphics[width=1\linewidth]{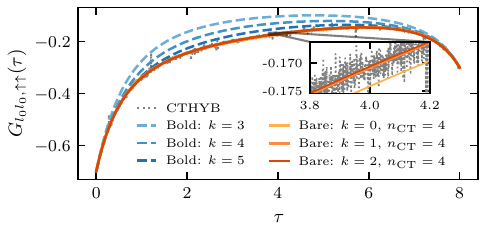}
    \caption{Two-orbital impurity model with a Slater–Kanamori interaction coupled to a semi-circular bath at  $\beta t = 8$. Plotted is the diagonal component of the Green’s function. $n_{\mathrm{CT}}:$ number of counterterms, $k:$ maximum expansion order. Bold data are reproduced from Fig.~6 of Ref.~\cite{huangAutomatedEvaluationImaginary2025}.}
    \label{fig:two_orbital_spinful_diag}
\end{figure}
We now consider an interacting two-orbital impurity model with a local Hamiltonian of the Kanamori type~\cite{kanamoriElectronCorrelationFerromagnetism1963}. Hamiltonians of this type are often used to describe Hund's metal physics~\cite{wernerSpinFreezingTransition2008,georgesStrongCorrelationsHunds2013}:
\begin{align}
\hat{H}_{\mathrm{loc}} 
&= U \sum_{l\in\{l_{0},l_{1}\}} \hat{n}_{l \uparrow} \hat{n}_{l \downarrow} +\hspace{-0.5cm} \sum_{\sigma, \sigma^{\prime} \in\{\uparrow, \downarrow\}}\left(U^{\prime}-J_H \delta_{\sigma \sigma^{\prime}}\right) \hat{n}_{l_{0} \sigma} \hat{n}_{l_{1} \sigma^{\prime}} \nonumber \\
&\quad + J_H\left(\hat{c}^{\dagger}_{l_{0} \uparrow} \hat{c}^{\dagger}_{l_{0} \downarrow} \hat{c}_{l_{1} \downarrow} \hat{c}_{l_{1} \uparrow}+\hat{c}^{\dagger}_{l_{0} \uparrow} \hat{c}^{\dagger}_{l_{1} \downarrow} \hat{c}_{l_{0} \downarrow} \hat{c}_{l_{1} \uparrow}+\text{h.c.}\right),\nonumber
\end{align}
where $\hat{n}_{l_{i}\sigma_{i}}\coloneqq \hat{c}^{\dagger}_{l_{i}\sigma_{i}}\hat{c}_{l_{i}\sigma_{i}}$ denotes the density operator. The first term describes the intra-orbital Coulomb repulsion, for which we choose $U=2$. The second term encodes the inter-orbital density–density interaction for antiparallel ($\sigma \neq \sigma'$) and parallel ($\sigma = \sigma'$) spins, with $U' = U - 2J_H$ in a rotationally invariant system and the Hund's coupling fixed to $J_H = 0.2$. The last term accounts for the pair-hopping and spin-exchange processes.
The chemical potential is set to $\mu = (3U - 5J_H - 3)/2 = 1$ and the inverse temperature to $\beta t = 8$. We consider a continuous bath with a semicircular density of states as before and take both diagonal and off-diagonal components of the hybridization function to be the same function, i.e., $\Delta_{l_i l_j,\sigma\sigma'}(\tau)
=
\delta_{\sigma\sigma'}\Delta(\tau)$ for $i,j\in\{0,1\}$.
Such off-diagonal hybridization functions induce a severe fermion sign problem in \ac{QMC} simulations based on the bare hybridization expansion~\cite{wernerContinuoustimeSolverQuantum2006,gullContinuoustimeMonteCarlo2011,hauleQuantumMonteCarlo2007,eidelsteinMultiorbitalQuantumImpurity2020} that can be remedied by using inchworm methods \cite{eidelsteinMultiorbitalQuantumImpurity2020}.

In Fig.~\ref{fig:two_orbital_spinful_diag}, we show that the diagonal component of the Green's function, computed using the bare expansion with $n_{\mathrm{CT}}=4$ (solid lines), converges rapidly as a function of the maximum expansion order $k$ toward the benchmark results (dotted) obtained from \ac{CTHYB} \cite{eidelsteinMultiorbitalQuantumImpurity2020}. For comparison, we also reproduce the bold expansion results of Ref.~\cite{huangAutomatedEvaluationImaginary2025} for the same model and parameters but without counterterms (dashed lines), which exhibit substantially slower convergence. A similar result is observed for the off-diagonal component, which is presented in the Supplemental Material~\cite{supplement_counterterms}.

The reported computational cost for the bold calculation of this model in Ref.~\cite{huangAutomatedEvaluationImaginary2025} is approximately $750$ core-hours for $k=5$. The inchworm \ac{QMC} calculation for the same problem presented in Ref.~\cite{eidelsteinMultiorbitalQuantumImpurity2020} required approximately $1500$ core-hours. The tensor-train approach in Ref.~\cite{yuInchwormTensorTrain2025}, applied to a similar problem with a discretized bath, requires a cost of roughly $500$ core-hours for each Green's function point. In contrast, the implementation of the bare  method combined with the counterterm technique required  $0.89$ core-hours for $k=2$ and $n_{\mathrm{CT}}=4$.

\begin{figure}[tb]
    \centering
    \includegraphics[width=1\linewidth]{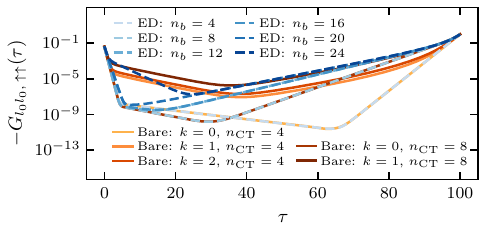}
    \caption{NiO \ac{SEET} impurity result (first iteration) at
    \(\beta=100\mathrm{Ha}^{-1}\). Shown is the diagonal
    part of the imaginary-time Green's function
    \(-G_{l_0l_0,\uparrow\uparrow}(\tau)\) on a logarithmic axis for the first impurity
    orbital \(l_0\), using the active-space setting \(a\) of
    Ref.~\cite{iskakovInitioSelfenergyEmbedding2020}. ED
     with \(n_b\) fitted bath states and bare solver with
     \(n_{\mathrm{CT}}\) counterterms at expansion order 
    \(k\).}
    \label{fig:NiO_up}
\end{figure}
Our final example is taken from the \ac{SEET} \cite{kananenkaSystematicallyImprovableMultiscale2015,zgidFiniteTemperatureQuantum2017,rusakovSelfEnergyEmbeddingTheory2019} NiO embedding construction discussed in
Ref.~\cite{iskakovInitioSelfenergyEmbedding2020} and performed at an inverse temperature \(\beta=100\mathrm{Ha}^{-1}\). The weakly correlated part is treated by self-consistent GW \cite{yehFullySelfconsistentFinitetemperature2022,iskakovGreenWeakCouplingImplementation2025}. We adopt active-space
choice \(a\) from Table~I in Ref.~\cite{iskakovInitioSelfenergyEmbedding2020},
consisting of two independent two-orbital impurity problems formed from the Ni \(e_g\) orbitals on the two antiferromagnetically inequivalent Ni sites. The comparison is carried out for the first \ac{SEET} iteration only, which is representative of impurity problems in this system and allows the exclusion of error propagation in the self-consistency cycle.
As a reference, we use a Lanczos \ac{ED} impurity solver~\cite{iskakovExactDiagonalizationLibrary2018,gaenkoUpdatedCoreLibraries2017}. The ED bath is obtained using
the ESPRIT-based hybridization fitting, which is more accurate than the non-linear fit of Ref.~\cite{iskakovInitioSelfenergyEmbedding2020}. The ED solver result is equivalent to the zeroth-order bare hybridization expansion result ($k=0$) for \(n_b\) equal to \(n_{\mathrm{CT}}\).

Figure~\ref{fig:NiO_up} shows the diagonal
component of the Green's function
\(-G_{l_0l_0,\sigma\sigma}(\tau)\) for one representative impurity
orbital with \(\sigma=\uparrow\) (see Supplemental Material~\cite{supplement_counterterms} for other components). 
ED shows a slow convergence from a strongly insulating solution,
characterized by a rapid decay of \(G(\tau)\), toward a solution with a smaller gap.
In our method, as either the maximum expansion order \(k\)
or the number of counterterms \(n_{\mathrm{CT}}\) is increased, the result converges to a solution with a smaller gap much more rapidly, though convergence cannot be established to the standards of the model systems. The example suggests that increasing the expansion order or using bold diagrammatics is necessary for full convergence.

In conclusion, we have introduced a hybrid Hamiltonian-diagrammatic quantum impurity solver in which a small set of 
counterterms obtained from an imaginary-time pole decomposition of the      
hybridization shifts the dominant bath contributions
to the local problem, leaving a small residual hybridization that can then be treated perturbatively.        
                                                    
For a simple test model, the method reaches $\varepsilon \sim 10^{-9}$,
surpassing the precision of \ac{MC} by many orders of magnitude. For a strongly       
correlated two-orbital model with a severe sign problem, rapid convergence is achieved at substantially decreased computational cost. For a realistic NiO embedding  
problem, results improve systematically with control parameters.
The counterterm framework can be combined with the full range of diagrammatic methods.
Replacing the bare expansion with bold-line~\cite{gullBoldlineDiagrammaticMonte2010,huangAutomatedEvaluationImaginary2025} or                                       
inchworm~\cite{cohenTamingDynamicalSign2015,eidelsteinMultiorbitalQuantumImpurity2020} summation may further accelerate convergence in realistic systems and substantially broaden the class of quantum impurity problems amenable to high-precision solution.

\begin{acknowledgments}
This material is based upon work supported by the US National Science Foundation under Grant No 2401159, which supported YY and EG. GH and DZ were supported by US NSF grant 2310582, LZ by US NSF grant 2310182, and XD by National Natural Science Foundation of China under grant No.12504289. 
We acknowledge fruitful discussions with Andr\'e Erpenbeck. The bare hybridization expansion solver used the \texttt{Lehmann.jl} package~\cite{kayeDiscreteLehmannRepresentation2022,kayeLibdlrEfficientImaginary2022,kayeCppdlrImaginaryTime2024} and the sum-of-exponentials technique introduced in Refs.~\cite{kayeDecomposingImaginaryTimeFeynman2024,huangAutomatedEvaluationImaginary2025}.\end{acknowledgments}

\textit{Data Availability.} All raw data of this study are available at \cite{yu_2026_20150672}.

\bibliography{counterterm_hyb}

\end{document}